\documentclass[twocolumn]{bmcart}

\usepackage{amsthm,amsmath}
\usepackage{amsfonts}

\usepackage{lipsum,url,booktabs,tabularx,multirow}
\usepackage{graphicx}
\usepackage{amsmath,amssymb}
\usepackage{chngcntr}

\usepackage[figuresright]{rotating}

\usepackage[utf8]{inputenc} 

\startlocaldefs
\endlocaldefs

\begin{document}

\begin{frontmatter}

\begin{fmbox}
\dochead{Research}


\title{Cost-effective Land Cover Classification for Remote Sensing Images}


\author[
  addressref={aff1,aff2},                   
  email={dongwei@bit.edu.cn, dli@swin.edu.au}   
]{\inits{D.L.}\fnm{Dongwei} \snm{Li}}

\author[
  addressref={aff1},
  email={slwang2011@bit.edu.cn}
]{\inits{S.W.}\fnm{Shuliang} \snm{Wang}}

\author[
  addressref={aff2},
  email={qhe@swin.edu.au}
]{\inits{Q.H.}\fnm{Qiang} \snm{He}}

\author[
  addressref={aff2},
  email={yyang@swin.edu.au},
  corref={aff2}
]{\inits{Y.Y.}\fnm{Yun} \snm{Yang}}


\address[id=aff1]{
  \orgdiv{School of Computer Science},             
  \orgname{Beijing Institute of Technology},          
  \city{Beijing},                              
  \cny{China}                                    
}
\address[id=aff2]{%
  \orgdiv{School of Software and Electrical Engineering},
  \orgname{Swinburne University of Technology},
  \city{Melbourne},
  \cny{Australia}
}



\end{fmbox}


\begin{abstractbox}

\begin{abstract} 
Land cover maps are of vital importance to various fields such as land use policy development, ecosystem services, urban planning and agriculture monitoring, which are mainly generated from remote sensing image classification techniques. Traditional land cover classification usually needs tremendous computational resources, which often becomes a huge burden to the remote sensing community. Undoubtedly cloud computing is one of the best choices for land cover classification, however, if not managed properly, the computation cost on the cloud could be surprisingly high. Recently, cutting the unnecessary computation \textit{long tail} has become a promising solution for saving cost in the cloud.  For land cover classification, it is generally not necessary to achieve the best accuracy and 85\% can be regarded as a reliable land cover classification. Therefore, in this paper, we propose a framework for cost-effective remote sensing classification. Given the desired accuracy, the clustering algorithm can stop early for cost-saving whilst achieving sufficient accuracy for land cover image classification. Experimental results show that achieving 85\%-99.9\% accuracy needs only 27.34\%-60.83\% of the total cloud computation cost for achieving a 100\% accuracy. To put it into perspective, for the US land cover classification example, the proposed approach can save over \$1,593,490.18  for the government in each single-use when the desired accuracy is 90\%.
\end{abstract}




\end{abstractbox}
%

\end{frontmatter}




\section*{Introduction}
Land cover maps represent the spatial information of different categories of physical coverage (e.g., forests, wetlands, grasslands, lakes, etc.) on surfaces of the earth \cite{landcover}, where dynamic land cover maps may contain changes in land cover categories over time, thereby capturing the changes of land arrangements, human activities, and the inputs people make within a land cover type to produce, alter or maintain it. Frequently updated land cover map is essential for a variety of environmental and socioeconomic applications, including urban planning \cite{bechtel2017beyond}, agricultural monitoring \cite{alcantara2012mapping}, forestry \cite{asner2006condition}, sustainable development \cite{glinskis2019quantifying} etc. 

Considering  the large geographic area and high temporal frequency covered by remote sensing satellite imagery, it provides a unique opportunity to obtain land cover information through the image classification process. Land cover classification is the grouping of pixels in the images into homogeneous regions, each of which corresponds to a specific land cover type, usually modelled as a clustering problem \cite{bensaid1996validity,zhang2016spectral}. Generally, unsupervised clustering is widely used in the land cover classification problem \cite{lu2007survey} because remote sensing images are often not available with ground truth of labels.

To generate updated land cover information at different scales, a series of remote sensing image classification techniques have been proposed in recent years \cite{li2014review}. Most representative clustering algorithms (e.g., \textit{k-means} \cite{celik2009unsupervised}, \textit{ISODATA} \cite{venkateswarlu1992fast}, \textit{Expectation-Maximum} \cite{kersten2005unsupervised}, \textit{Markov Random Field} (MRF) \cite{xu2013unsupervised}) consider the pixel as the basic analysis unit, with which each pixel is labeled as a single land cover type. However, these pixel-wise clustering approaches, when applied to heterogeneous regions, may have limitations as the size of an object may be much smaller than the size of a pixel. In particular, a pixel may not only contain a single land cover type, but a mixture of several land cover types. Therefore, fuzzy clustering approaches have been developed for unsupervised land type classification \cite{wang2013adaptive, sowmya2011land}.   

The advancement of spatial, spectral, temporal and angular data has facilitated the generation of petabytes of data every year \cite{gao2022generative,kjaergaard2020current,gao2021understanding}. Land cover classification usually needs tremendous computational resources and becomes a huge burden to the remote sensing companies and organisations. With the ever-increasing demand for storing and analyzing large volumes of remote sensing imagery, cloud computing offers a suitable solution for the remote sensing community  \cite{fu2018secure}. By acting as a near-real-time insight platform, cloud computing can rapidly perform big data analysis. It is a mature platform that provides global users with high-end computing resources without a huge IT infrastructure investment budget, and provide efficient and low-cost solutions for remote sensing classification.

\begin{figure}[h!]
    \centering
    \includegraphics[width=0.9\linewidth]{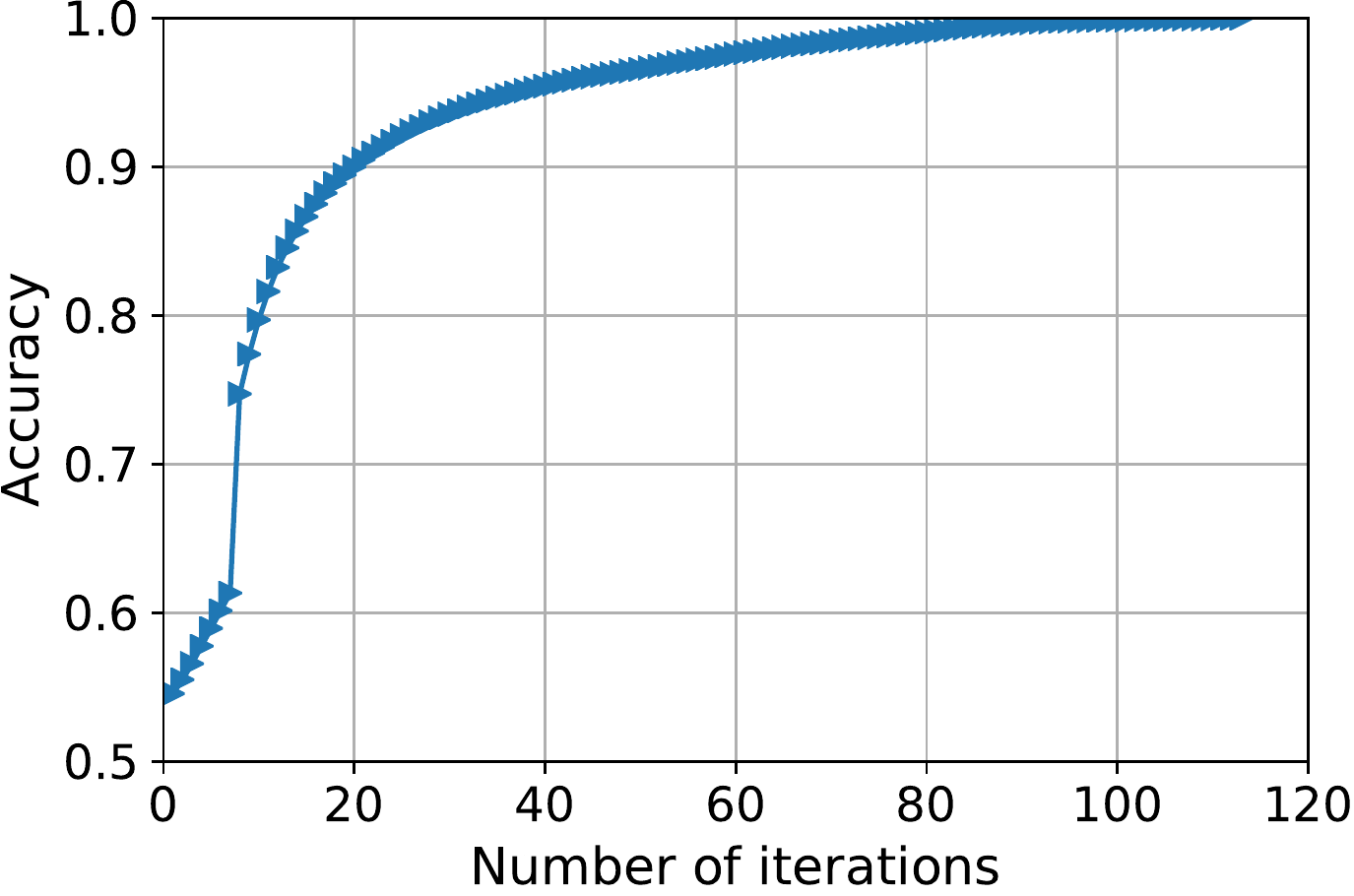}
    \caption{Long tail phenomenon during the clustering process}
    \label{fig:longtail}
\end{figure}

However, the cost of cloud computing environments for big data storage and analytics is drawing increasing attention from researchers, which becomes a bottleneck for land cover classification in the cloud. For example, running 100 m4-2xlarge EC2 virtual machines (VM) instances in Amazon Sydney datacenter costs up to \$62,496 per month \cite{Amazon}. Li et al. \cite{li2019cutting} found that cutting the unnecessary \textit{long tail} (see Fig.~\ref{fig:longtail}) in the clustering process is a promising solution for cost-effective cloud computing, which inspires us that we can explore achieving sufficiently satisfactory clustering accuracy with the lowest possible computation cost. In particular, this method could be effectively applied to the land cover classification.

In most clustering scenarios (e.g., spatial data analysis, weather forecast, marketing), we do not always need to have the best solution because users usually don't need 100\% accuracy. Taking weather forecast as an example, clustering techniques have been used to predict weather conditions (e.g., rainy, snowy, sunny) based on various factors such as air temperature, air pressure, humidity, amount of cloud cover, and speed of the wind. In this case, a reasonable margin of error is acceptable because users do not need to know 100\% accurate weather information. As long as they have a general understanding of the weather, they will be able to make decisions about what to wear or whether to bring an umbrella when going out. In the real world, there will never be completely accurate for clustering, such as weather forecasting and land cover classification. It is critical to stop clustering at a reasonable point to save computation costs if achieving a sufficiently satisfactory accuracy at a low cost is preferable to achieving 100\% accuracy at a high cost.

For the land cover classification problem, it is also not necessary to achieve the best accuracy all the time. Normally at least 85\% accuracy can be considered a reliable land cover classification \cite{anderson1976land}. To achieve cost-effective land cover classification, a new framework needs to be explored to improve  cost-effectiveness performance, rather than using the same methods in the general big data clustering scenarios. In general, there are three main challenges to be addressed for the design of the new framework: 1) unlike traditional pixel-wise clustering methods, we should adopt fuzzy clustering methods (e.g., the FCM algorithm) to assign pixels to multiple land cover types; 2) before building the regression models between the change rate of objective function and accuracy, we should first detect and remove the anomalies; 3) compared to the commonly used quadratic polynomial regression in previous literature \cite{li2019cutting}, more regression models should be explored for more cost-effective land cover classification.

In this research, we propose a generalized framework for cost-effective land cover classification with remote sensing images. We are the first to apply the FCM clustering algorithm to cost-effective land cover classification. \textit{Rand Index} is used as the accuracy metric and \textit{Local Outlier Factors} (LOF) \cite{breunig2000lof} is employed to remove anomalies between the change rate of objective function and accuracy. \textit{Support Vector Regressor} (SVR) \cite{cawley2004fast} is applied to fit the relation between the change rate of objective function and accuracy.  Experimental results show that achieving 85\%, 90\%, 95\%, 99\%, 99.9\% accuracy need only 27.34\%, 29.33\%, 33.25\%, 55.93\% and 60.83\% of the computation cost required for achieving a 100\% accuracy.  Our contributions are as follows:

\begin{itemize}
    \item We propose a generalized framework for the cost-effective land cover classification problem, with which the clustering algorithm can stop at an early point given the desired accuracy for cost-saving.
    \item We are the first to adopt the LOF algorithm to remove anomalies before fitting the relation between the change rate of objective function and accuracy, which improves the cost-effectiveness in the land cover classification.
    \item Experimental results show that the proposed framework can achieve sufficient accuracy and save much computation cost in the cloud.
\end{itemize}

The remainder of the paper is organised as follows. Section 2 discusses the current related works on remote sensing classification and the cost of cloud computing. In Section 3, we introduce the background knowledge used in our study and in Section 4 we demonstrate our generalized framework for land cover classification. Then, in Section 5, we conduct extensive experiments to illustrate the cost-effectiveness of the proposed framework. Section 6 gives conclusions and future work.

\section*{Related Works}
\label{sec:relatedwork}
\subsection*{Remote Sensing Imagery Classification}
Fuzzy C-means (FCM) is first proposed by Dunn and improved by Bezdek \cite{bezdek2013pattern}, which is frequently used in the image segmentation field. Foody et al. \cite{zhang1998fuzzy} used the FCM algorithm for sub-urban land use mapping from remote sensing images. They found that the classification results can be improved significantly when using fuzzy clustering compared with hard clustering methods.

Wang et al. \cite{wang2013adaptive} incorporated the spatial context to improve the robustness of the FCM algorithm in image segmentation. By combining these two concepts and modifying the objective function of the FCM algorithm, they solved the problems of sensitivity to noisy data and the lack of spatial information, and improved the image segmentation results. Sowmya et al. \cite{sowmya2011land} proposed the reformed fuzzy C-means (RFCM) technique for land cover classification. Image quality metrics such as error image, peak signal to noise ratio (PSNR) and compression ratio were used to compare the segmented images.

\subsection*{Cost-effective Cloud Computing}
With the development of the pay-as-you-go cost model, IT resources are often provided and utilized by cloud computing. Since most of the benefits offered by cloud computing are around the flexibility of the pay-as-you-go model, cost-effectiveness has become a key issue in the cloud computing area. With the continuous improvement of cloud services provided by cloud vendors, many scientists are beginning to pay attention to the performance and cost-effectiveness of public cloud services. In-depth research has been conducted on cost-effective computation in cloud environments. 

Cui et al. \cite{cui2019tailcutter} identified the high tail latency problem in cloud CDN via analyzing a large-scale dataset collected from 783,944 users in a major cloud CDN. A workload scheduling mechanism was presented aiming to optimize the tail latency while meeting the cost constraint given by application providers. A portfolio optimization approach was then proposed by \cite{alam2019edge} for cost-effective healthcare data provisioning. Li et al. \cite{li2019performance} modelled the task scheduling on IoT-cloud infrastructure as bipartite graph matching, and proposed a resource estimating method.

A semi-elastic cluster computing model \cite{niu2016building} was introduced for organizations to reserve and dynamically adjust the size of cloud-based virtual clusters. The experiment results indicated that such a model can save more than 60\% cost for individual users acquiring and managing cloud resources without leading to longer average job wait times. The MapReduce cloud model Cura was proposed to offer a cost-effective solution to effectively deal with production resources, which implemented a globally effective resource allocation process that significantly reduces the cost of resource use in the cloud. Flutter \cite{hu2018time} was designed and implemented as a task scheduler and reduced the completion time and the network cost for large-scale data processing tasks over data centres in different regions.  
 
Berriman et al. \cite{berriman2010application} used Amazon EC2 to study the cost-effectiveness of cloud computing applications and Amazon EC2 was compared with the Abe high-performance cluster. They concluded that Amazon EC2 can provide better performance for memory- and processor-bound applications than I/O-bound applications. Similarly, Carlyle et al.\cite{carlyle2010cost} compared the computation cost of high-performance in Amazon EC2 environments and traditional HPC environments with Purdue University's HPC cluster program. Their research indicated that the in-house cluster can be more cost-effective while organizations take advantage of clusters or have IT departments that can maintain an IT infrastructure or prioritize cyber-enabled research. These features of in-house clusters actually demonstrated the cost-effectiveness and flexibility of running computation-intensive applications in the cloud. 

A random multi-tenant framework was proposed by Wang et al. \cite{wang2017optimizing} for investigating the cloud services response time as an indicator with a universal probability distribution. Similarly, Hwang et al. \cite{hwang2016cloud} tested the performance of Amazon cloud services with 5 different benchmark applications and found it was more cost-effective in sustaining heavier workload, by comparing the scale-out strategies and the scale-up strategies. To explore the minimal cost of storing and regenerating data sets in multiple clouds, \cite{yuan2018algorithm} proposed a novel algorithm that implements the best compromise among storage, bandwidth, and computation cost in the cloud. Jawad et al. \cite{jawad2018robust} proposed an intelligent power management system in order to minimize data centre operating costs. The system can coordinate the workload of data centre, renewable power, battery bank, diesel generators, real-time transaction price for the purpose of reducing the cost of consumption. Aujla et al. \cite{aujla2017optimal} proposed an efficient workload slicing scheme for handling data-intensive applications in multi-edge cloud environments using software-defined networks (SDN) to reduce the migration delay and cost.

The current research on cloud computing indicates the prevalence of running computation-intensive applications in the cloud, which provides a general overview of the cost-effectiveness of big data analysis in the cloud by comparing traditional cluster environments and cloud environments. In order to save costs in the cloud, it is also important for algorithms to reduce processing time and improve their efficiency. Li et al. \cite{li2019cutting} proposed a method for cutting the unnecessary long tail in the clustering process to achieve cost-effective big data clustering in the cloud. Sufficiently satisfactory accuracies can be achieved at the lowest possible costs by setting the desired accuracies, which presented an important step toward cost-effective big data clustering in the cloud. In this research, we adopt the approach proposed in \cite{li2019cutting} to a more specific field: remote sensing land cover classification, and explore more advanced and efficient ways to improve the performance of cost-effective clustering in the cloud.

\section*{Background}
This section mainly introduces the background of the proposed cost-effective land cover classification method, including the fuzzy c-means clustering algorithm, accuracy calculation method, and the cloud cost computing model. 

\subsection*{Fuzzy C-means Clustering}
As one of the most commonly used fuzzy clustering methods, the FCM algorithm \cite{bezdek1984fcm,bezdek2013pattern} is a clustering technique allowing each data point to belong to more than one cluster. Fuzzy logic principles are used to assign each point a membership in each cluster center from 0 to 1, which indicates the degree to which data points belong to each cluster. Therefore, the FCM algorithm can be very powerful compared to traditional hard clustering (i.e., K-means \cite{hartigan1979algorithm}) where every point can only belong to exactly one class. 
FCM clustering is based on minimizing the objective function as follows:
\begin{equation}
    \mathcal{J}_m = \sum_{i=1}^{N}\sum_{j=1}^{C} 
    u_{ij}^m {\left \| x_i - c_j \right \|}^2,     1\leq m< \infty,  m \in \mathbb{R}
\end{equation}
where $m$ is a real number larger than 1 and means the $m$th iteration during the clustering process. $u_{ij}$ means the degree of membership of $x_i$ in the cluster $j$, $x_i$ indicates the $i$th d-dimensional measured data, $c_j$ is the $j$th d-dimension center of the cluster, and $\left \| * \right \|$ is any norm expressing the similarity between any measured data and the center. Fuzzy partitioning is conducted through an iterative optimization of the objective function shown below, with the update of membership $u_{ij}$ and the cluster centers $c_j$ by:
\begin{equation}
u_{ij} = \frac{1}{{\sum_{k=1}^{C}}{\left (\frac{\left \| {x_i - c_j}  \right \|}{\left \| {x_i - c_k}  \right \|}  \right )}^{\frac{2}{m-1}}}
\end{equation}

\begin{equation}
    c_j = \frac{\sum_{i=1}^{N}u_{ij}^m\cdot x_i}{\sum_{i=1}^N u_{ij}^m}
\label{equ: c_j}
\end{equation}

This iteration will stop when \begin{equation}
    {max}_{ij} \left \{ \left | u_{ij}^{k+1} - u_{ij}^k \right | \right \} < \varepsilon,  0<\varepsilon < 1,  \varepsilon \in \mathbb{R}
\end{equation}
where $\varepsilon$ is a termination criterion between 0 and 1, whereas $k$ is the iteration step. This procedure converges to a local minimum or a saddle point of $ \mathcal{J}_m$. Overall, the algorithm is composed of the following steps:
\begin{enumerate}
    \item Initialize matrix $U= [u_{ij}] $ as $U^{0}$.
    \item In $k$ step, calculate the centers vectors $C^k = [c_{ij}]$ with $U^k$ based on Equation (\ref{equ: c_j}).
    \item Update $U^k$ and $U^{k+1}$.
    \item If $\left \| U^{k+1} - U^{k} \right \| < \varepsilon$, then stop; other wise, return to Step 2.
\end{enumerate}

\subsection*{Rand Index}

Accuracy is a key metric for assessing the effectiveness of big data clustering. As suggested by \cite{li2019cutting}, to demonstrate that the clustering accuracy gradually increases iteration by iteration, we adopt the final clustering partition  $P_f$ as a reference partition as 100\% accuracy. By comparing the clustering results obtained in each iteration, we exhibit how the accuracy of the intermediate partition $P_i \in \{P_1,P_2,...,P_f \}$ increases during the clustering process.

In our research, the accuracy of the clustering algorithm can be measured by the similarity between $P_i$ and $P_f$. \textit{Rand Index} \cite{rand1971objective} is adopted to evaluate the similarity between two clustering partitions, which is a popular method of accuracy calculation in the field of data clustering. Each partition is treated as a group of $(m-1) \times m/{2}$ pairs of data points, where $m$ represents the size of the dataset. For each pair of data points, the partition either assign them to the same cluster or different clusters. Therefore, the similarity between the partitions $P_1$ and $P_2$ can be measured as follows:
\begin{equation}
\label{equ:rand}
    Rand(P_1,P_2) = \frac{m_{00}+m_{11}}{m_{00}+m_{01}+m_{10}+m_{11}} = \frac{m_{00}+m_{11}}{\binom{m}{2}}
\end{equation} 
where:

$m_{00}$ indicates the number of data point pairs located in the different clusters in both $P_1$ and $P_2$;

$m_{11}$ indicates the number of data point pairs located in the same clusters in both $P_1$ and $P_2$;

$m_{01}$ indicates the number of data point pairs located in the same clusters in $P_1$ but in different clusters in $P_2$;

$m_{10}$ indicates the number of data point pairs located in different clusters in $P_1$, but in the same clusters in $P_2$.

\begin{figure}
    \centering
    \includegraphics[width=0.85\linewidth]{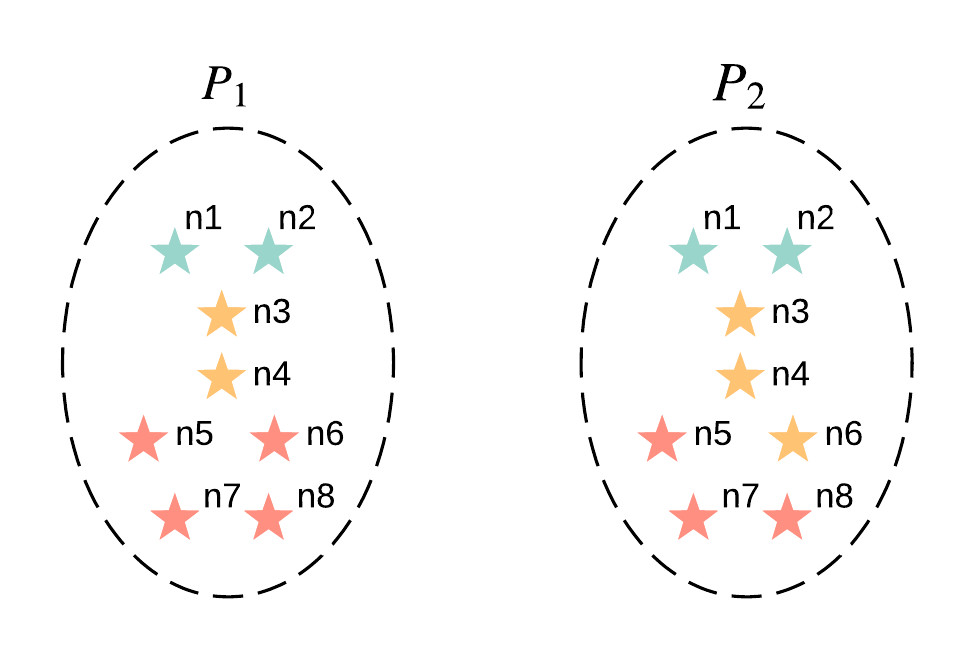}
    \caption{An example of calculating \textit{Rand Index} between $P_1$ and $P_2$}
    \label{randindex}
\end{figure}
 
With the \textit{Rand Index} as the measure of similarity, the clustering accuracy can be calculated in each iteration of the clustering process. Take Fig.~\ref{randindex} for instance, the data point pairs located in the same cluster (indicated with same color) in $P_1$ and $P_2$ includes $(n1, n2)$, $(n3, n4)$, $(n5, n7)$, $(n5, n8)$, $(n7, n8)$. The pairs that are placed in different clusters in both $P_1$ and $P_2$ include $(n1, n3)$, $(n1, n4)$, $(n1, n5)$, $(n1, n6)$, $(n1, n7)$, $(n1, n8)$, $(n2, n3)$,$(n2, n4)$, $(n2, n5)$, $(n2, n6)$, $(n2, n7)$, $(n2, n8)$, $(n3, n5)$, $(n3, n7)$, $(n3, n8)$, $(n4, n5)$, $(n4, n7)$, $(n4, n8)$. Then, there is $Rand(P_1,P_2 )=(5+18)/28=82.14\%$. Clearly, the value of \textit{Rand Index} increases as the number of iterations increases. In the last iteration of clustering process where $P_i=P_f$, there is $Rand(P_i,P_f )=1$, indicating that the clustering process completes with a 100\% accuracy.

\subsection*{Cloud Computing Model} 
The computation cost for remote sensing image classification can be computed by the cost models provided by cloud vendors. Amazon EC2 web services are adopted in this research which usually have 4 different models: spot instances, on-demand, dedicated hosts and reserved instances. As the most basic cost model, on-demand model is paid by time and does not require upfront payments or long-term commitments. Therefore, the on-demand cost model is adopted in this research for calculating the computation cost in the cloud.
\begin{equation}
    \textrm{Cost}_\textrm{comp} =\textrm{Price}_\textrm{unit} \times {T}_\textrm{comp}
\end{equation}
Similar to \cite{li2019cutting}, computation time $T_\textrm{comp}$ is calculated with the time taken during the clustering process. The unit price $\textrm{Price}_\textrm{unit}$ is defined by the computational resource used in running the algorithm. Take Amazon EC2 for instance, there are 7 major types of EC2 virtual machine instances: RHEL, SLES, Linux,  windows,  Windows with SQL Web, Windows with SQL Enterprise and windows with SQL Standard. Different types of EC2 VM instances have different unit prices. For example, in Windows type, 36 EC2 VM instances are displayed for 4 types: Compute Optimized, General Purpose, Storage Optimized and Memory Optimized. Unit prices vary from region to region, ranging from \$0.0066 to \$38.054 per hour. 

In this paper, for the sake of simplicity, the computation time is used as an indicator for calculating the computation cost. When we use a specific Amazon EC2 VM instance, we can see that the computation time and computation cost are positively correlated. The longer the computation time, the higher the computation cost. Some other costs may occur before running the algorithm, such as data transfer costs and storage costs for large data sets in the cloud. However, the cost of data storage and data transfer is independent of the clustering process. Therefore, in this study, we only focus on the computational cost of the land cover classification process and isolate it from other costs.

\section*{Approach}
\label{sec:framework}

\begin{figure*}
    \centering
    \includegraphics[width=0.85\textwidth]{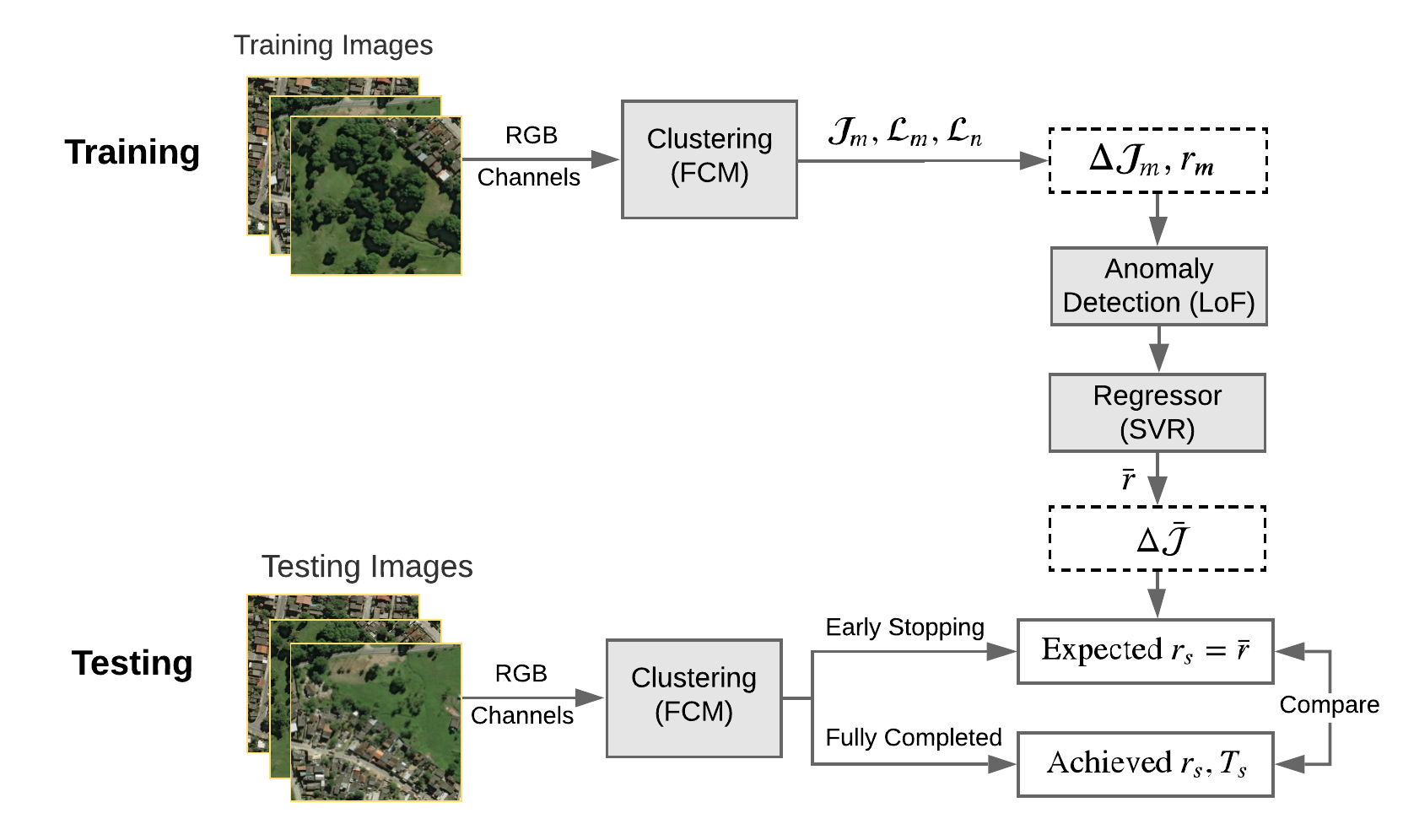}
    \caption{Framework of cost-effective remote sensing image classification}
    \label{fig:framework}
\end{figure*}

Fig.~\ref{fig:framework} shows the proposed framework consisting of two phases: training phase and testing phase. For the training phase, we learn the relation between the accuracy and the change rate of objective function. Through the testing phase, we set the desired accuracy and stop the clustering algorithm at an early point by meeting sufficient accuracy. The detailed process is as follows:

\subsection*{Training phase}

In the training phase, the FCM clustering algorithm is applied on RGB channels of training images. During the clustering process, we get $\mathcal{J}_m$ and $\mathcal{L}_m$, indicating the objective function and the predicted labels at the $m$th iteration of total $n$ iterations, where the predicted labels at the last iteration are noted with $\mathcal{L}_n$. Once the clustering is finished, $r_m$ (the accuracy at $m$th iteration) is calculated through the \textit{Rand Index} between $\mathcal{L}_m$ and $\mathcal{L}_n$ based on Equation (\ref{equ:rand}). 

The rate of change of objective function $\Delta \mathcal{J}_m$ is computed using the Equation (\ref{equ: delta}). For simplicity, we use 'change rate' instead of 'the rate of change' in this paper. The change rate is used to describe the percentage change in value over a specific period of time. In this research, we define the change rate of objective function as:
\begin{equation}
    \label{equ: delta}
    \Delta{\mathcal{J}_m} = \frac{\mathcal{J}_m}{\mathcal{J}_{m-1}}\times 100\%, 1<m\leqslant n
\end{equation}
where $\Delta \mathcal{J}_m$ indicates the change rate of objective function at the $m$th iteration of total $n$ iterations.

For each training image, we can calculate the value of $r_m$ and $\Delta \mathcal{J}_m,  m\in\{2,...n\}$. As a result, we get $n-1$ data points for each training image. To model the relationship between $r_m$ and $\Delta \mathcal{J}_m$, anomaly points need to be mitigated first. As the most well-known anomaly detection algorithm, LOF \cite{breunig2000lof} is an unsupervised machine learning algorithm that finds anomalies by measuring the local deviation of a given data point based on its neighbors. In our research, LoF algorithm is applied to mitigate the anomaly points. 

With anomalies removed, we have tried several commonly used regression models to fit the relation between $r_m$ and $\Delta \mathcal{J}_m$ in the remaining points, such as SVR, \textit{Standard Linear Regressor} (LR) \cite{seber2012linear}, \textit{Gradient Booting Regressor}(GBR) \cite{friedman2001greedy}, \textit{Bayesian Ridge Regressor} \cite{tipping2001sparse} and \textit{Random Forest Regressor} (RFR) \cite{liaw2002classification}. \textit{Support Vector Machine} (SVM) \cite{cawley2004fast} in regression problems, commonly known as SVR, is one of the most widely used regression models. LR is a linear model which assumes the linear relationship between two variables. GBR is an ensemble method that combines a set of weak predictors to achieve reliable and accurate regression. \textit{Bayesian Ridge Regressor} formulates linear regression by using probability distributions. RFR follows the idea of random forest in classification and can estimate the importance of different features. 

\begin{figure*}
    \centering
    \includegraphics[width=0.55\textwidth]{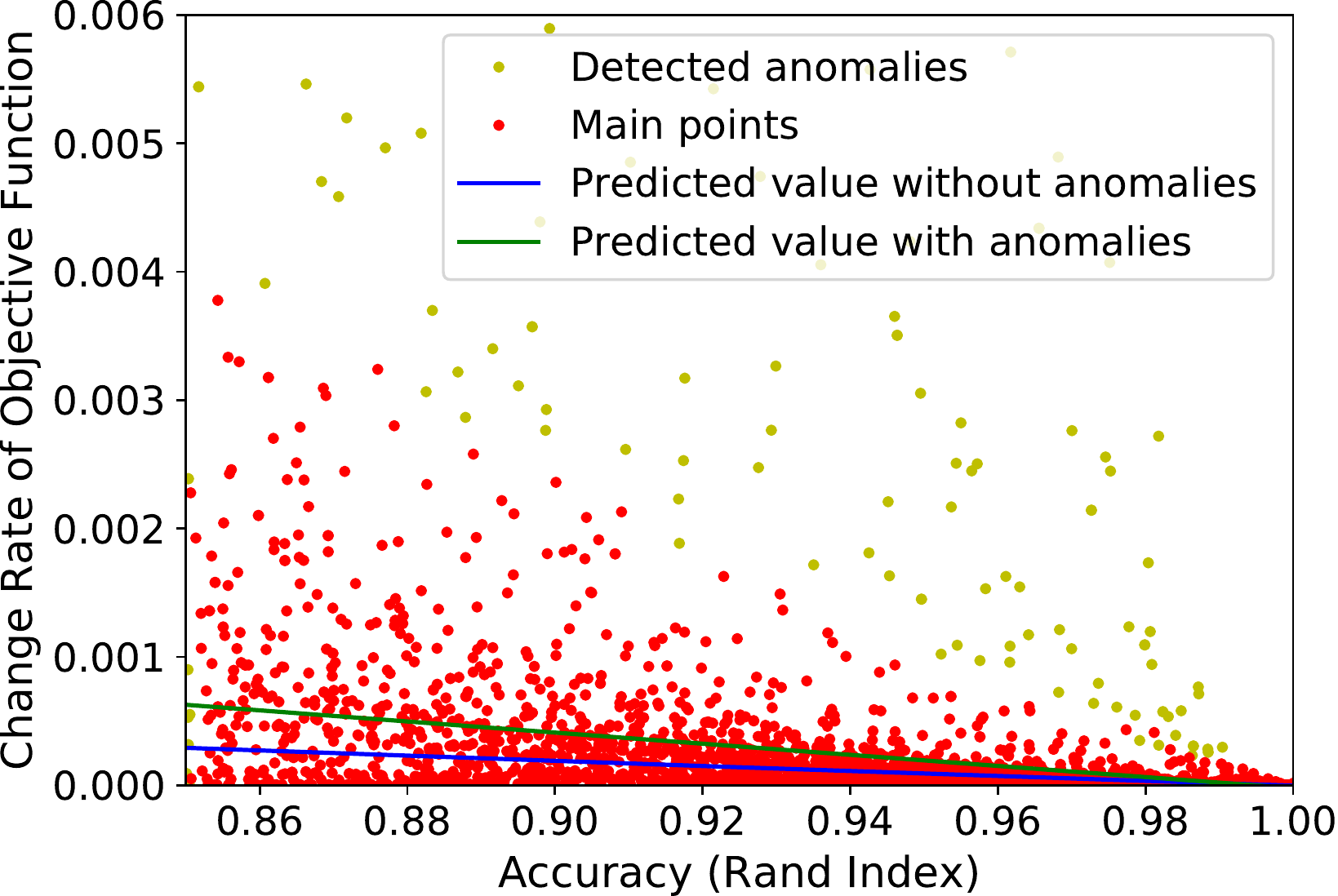}
    \caption{Relation between change rate of objective function and accuracy}
    \label{fig:points}
\end{figure*}

After extensive experiments, we found that when the SVR is applied, the experiment results usually show better performance. As a result, SVR is adopted as the regression model to fit the relation between $r_m$ and $\Delta \mathcal{J}_m$. Given the desired accuracy $\bar{r}$ (e.g., 85\%, 90\%, 95\%, 99\%, 99.9\%), the predicted value of $\Delta \bar{\mathcal{J}}$ can be calculated from the trained regressor (see Fig.~\ref{fig:points}).

\subsection*{Testing phase}
In the testing phase, we run the FCM clustering algorithm with the testing images. $\Delta\mathcal{J}$ at each iteration is calculated and compared with $\Delta \bar{\mathcal{J}}$. When $\Delta\mathcal{J}< \Delta\bar{\mathcal{J}} $, we record the early stopping point at this iteration, e.g., $s$th iteration. In the real scenario of remote sensing classification, we can stop the clustering algorithm at this point with the confidence of achieving the desired accuracy $\bar{r}$ at the $s$th iteration. 

\textbf{Evaluation Method}. To evaluate the performance of the proposed approach, we run the FCM algorithm until it is fully completed during training. Then we calculate the achieved accuracy ${r}_s$ and computation time $T_s$ at the $s$th iteration. Finally, we can evaluate the proposed method from two dimensions: the achieved accuracy (through comparing $r_s$ and $\bar{r}$) and the percentage of saved time ($T_s/T_n$).

\textbf{Cloud Computation Cost}. Total computation time ${T}_\textrm{comp}$ includes the overall clustering time in the training process ${T}_\textrm{train}$, and the actual clustering time  ${T}_\textrm{actual}$ (i.e., early-stop computation time) when clustering reaches the desired accuracy, which is computed as:

\begin{equation}
    {T}_\textrm{comp} = {T}_\textrm{train} + T_\textrm{actual}
\end{equation}

The training phase is carried out only once. Once it is completed, the regression model can be applied repeatedly to the remote sensing image classification in the future. Thus, ${T}_\textrm{train}$ can be negligible compared with the overall cost in the long term. Since the computation time is the primary indicator of the cost in our research, the cost-effectiveness percentage $\textrm{Cost}_\textrm{effective}$ can be exhibited as follows:

\begin{equation}
    \textrm{Cost}_\textrm{effective} \approx \frac{{T}_\textrm{actual}}{{T}_\textrm{total}}
    \label{costeffective}
\end{equation}
where ${T}_\textit{total}$ represents the expected computation time in the clustering when 100\% accuracy is achieved. The smaller the value of $\textit{Cost}_\textit{effective}$ is, the higher the cost-effectiveness of the clustering.

\section*{Experimental Evaluation}
In this section, we first introduce the experimental settings and the dataset. Then we conduct the experiments consisting of the training phase and testing phase. After that, we evaluate the proposed framework from two aspects: the achieved accuracy and the cost-effectiveness. Finally, we discuss the performance of the cost-effective land cover classification and real-world applications.

\subsection*{Experimental Setup}
The experiments were conducted on a laptop (Microsoft Corporation - Surface Laptop 4) with a 2.60 GHz Intel (R) Core (TM) i5 processor and 8G memory, and the operating system is 64-bit Windows 10 enterprise. The code was written in Python 3.6 and developed in PyCharm 4.5 IDE, making use of Scikit-learn, skfuzzy, Numpy, Pandas, SciPy and Matplotlib package for machine learning, mathematical, statistical operation and visualization. 

We conduct experiments on the public satellite imagery dataset SpaceNet \cite{Spacenet}. The dataset is released by Digital Globe, an American vendor of space imagery and geospatial content. The dataset includes a large amount of geospatial information related to various downstream use cases, e.g., infrastructure mapping and land cover classification. SpaceNet contains more than 17,533 high-resolution remote sensing images ($438 \times 406 $ pixels). SpaceNet is hosted as the Amazon Web Services public dataset, which contains approx. 67,000 square kilometers of high-resolution imagery in different cities (e.g., Las Vegas, Khartoum, Rio De Janeiro, Shanghai), more than 11 million building footprints, and approx. 20,000 kilometers of road labels, making it the most popular open-source dataset for geospatial machine learning research \cite{li2019cutting,van2021multi}. Due to the huge size of the SpaceNet data set, we randomly select 200 sample remote sensing images as the training data set so that we could perform the clustering and simulate the regression process accurately.
  
In the experiment, the FCM clustering algorithm (\textit{ncenters} = 6, \textit{error} = 0.005, \textit{m} = 2) was applied for cost-effective remote sensing image classification. Usually, finding the optimal number of clusters is crucial for the unsupervised clustering. For the SpaceNet dataset, through visual inspection, we find that the images can be generally divided into six different regions, i.e., forest, water, road, building, grassland and wasteland. Therefore, we set the number of clusters \textit{ncenters} = 6. The \textit{m} is an array exponentiation applied to the membership function at each iteration which is usually set to 2 for the FCM algorithm. The \textit{error} indicates the stopping criterion and we use the default value \textit{error} = 0.005 like previous studies \cite{li2019cutting}. 

After clustering, the LoF technique (\textit{outliers\_fraction} = 0.03,  \textit{n\_neighbors} = 40) was applied to remove the anomalies. For the parameters \textit{outliers\_fraction} and \textit{n\_neighbors}, we experimented with different parameter settings, and we achieve the best performance of the proposed method using the above settings. Next, SVR (kernel='RBF') was used to predict the change rate of objective function with the desired accuracy (i.e., 85\%, 90\%, 95\%, 99\%, 99.9\%). We choose the desired accuracy from 85\% because it is generally regarded as a reliable accuracy for land cover classification \cite{anderson1976land}. Then, we evaluate the proposed approach from two dimensions: the achieved accuracy and the percentage of saved time.

\begin{table*}
\caption{Change rate of objective function with different desired accuracies}
\label{tab:objective function}
\begin{tabular}{@{}llllll@{}}
\toprule
\multirow{2}{*}{\textbf{Algorithm}} & \multicolumn{5}{l}{\textbf{Desired accuracy}}                  \\ \cmidrule(l){2-6} 
                                                            & \textbf{85\%} &\textbf{90\%} & \textbf{95\%} & \textbf{99\%} & \textbf{99.9\%} \\ \cmidrule(r){1-6}
\textit{FCM}                                     &2.67e-4           & 1.76e-4       & 8.59e-5       & 1.61e-6       & 6.50e-7         \\
 \bottomrule
\end{tabular}
\end{table*}

\subsection*{Experiment Results}

Our experiment includes two phases: training phase and testing phase. For training remote sensing images, we first cluster the pixels in RGB channels with the FCM algorithm. We compute the objective function $\mathcal{J}_m$, predict label $L_m$ at the $m$th iteration until the last iteration $n$. Then, in each iteration, the change rate of objective function $\Delta{\mathcal{J}_m}$ is computed based on from Equation  (\ref{equ: delta}) and the accuracy $r_m$ is computed from Equation (\ref{equ:rand}). Fig.~\ref{fig:points} shows the relation between $\Delta{\mathcal{J}_m}$ and $r_m$.

After that, the LoF technique is used to remove the anomalies. In Fig.~\ref{fig:points}, red points represent the normal ones and yellow dots mean the detected dots anomalies. SVR is then applied to fit the relation between $r_m$ and $\Delta \mathcal{J}_m$. The green line represents the regression line with anomalies and blue line means the fitting line without anomalies. It can be observed that, given the same desired accuracy, the predicted value with anomalies (green line) is generally smaller than the predicted value without anomalies (blue line). 

Given the desired accuracy $\bar{r}$, we can predict the corresponding $\Delta \bar{\mathcal{J}}$. Table~\ref{tab:objective function} shows the different change rate of objective function with different required accuracies (e.g. 85\%, 90\%, 95\%, 99\%, 99.9\%) for the FCM clustering algorithm. The results show that, in the real-world scenario, if the desired accuracy is $\bar{r}$ (i.e., 99\%), we can apply the FCM algorithm on the remote sensing images, compute the change rate of objective function at each iteration and stop the algorithm when the change rate of objective function is below $\Delta\bar{\mathcal{J}}$ (e.g., 1.84e-6). However, when we make the FCM algorithm stop at an early point, is the achieved accuracy really up to 99\%? How much time could we save by this approach? To evaluate the performance of our method, we propose two criteria: the achieved accuracy and cost-effectiveness (the percentage of saved time).

\textbf{Achieved Accuracy}.
To evaluate our proposed method, given the desired accuracy of $\bar{r}$, we first calculate the corresponding change rate of objective function (see Table~\ref{tab:objective function}). Then, we run the FCM algorithm on testing images, calculate the change rate of objective function until it reaches $\Delta\bar{J}$ at the $s$th iteration. 

In this research, we complete the whole clustering process and calculate the achieved accuracy at the $s$th iteration. After that, the achieved accuracy $r_s$ is compared with the desired accuracy $\bar{r}$. Table~\ref{tab:averageaccuracy} shows the result of the average achieved accuracy (with standard deviation) of different desired accuracy for the FCM algorithm. We can see that the achieved accuracy is very close yet above the given desired accuracy and even higher than the desired accuracy. For example, on average, the achieved accuracy reaches 99.27\% when the desired accuracy is 99\%, and 99.92\% when the desired accuracy is 99.9\%. This illustrates that our proposed method has high accuracy on the FCM algorithm.

\begin{table*}
    \caption{Average achieved accuracy over different desired accuracies}
    \label{tab:averageaccuracy}
    \begin{tabular}{@{}llllll@{}}
    \toprule
    \multirow{2}{*}{\textbf{Aglorithm}} & \multicolumn{5}{l}{\textbf{Desired accuracy}}                                        \\ \cmidrule(l){2-6}    & \textbf{85\%} 
                                        & \textbf{90\%}                                              & \textbf{95\%}                                              & \textbf{99\%}                                              & \textbf{99.9\%}                                            \\ \cmidrule(r){1-6}
    \textit{FCM}                        & 
    \begin{tabular}[c]{@{}l@{}}89.15\%\\ (0.0594)\end{tabular} &
    \begin{tabular}[c]{@{}l@{}}93.16\%\\ (0.0544)\end{tabular} & \begin{tabular}[c]{@{}l@{}}95.07\%\\ (0.0370)\end{tabular} & \begin{tabular}[c]{@{}l@{}}99.27\%\\ (0.0044)\end{tabular} & \begin{tabular}[c]{@{}l@{}}99.92\%\\ (0.0018)\end{tabular} \\
    \bottomrule
    \end{tabular}
\end{table*}

\textbf{Cost-effectiveness}.
Table~\ref{tab:actual time} shows the actual percentage of saved computation time with different desired accuracy for the FCM algorithm. It can be found that we only use 27.34\%, 29.33\%, 33.25\%, 55.93\%, 60.83\% computation time when the desired accuracies are 85\%, 90\%, 95\%, 99\% or 99.9\% respectively. Since the cost of cloud computation is directly related to the actual computation time, the FCM algorithm can achieve high cost-effectiveness in the cloud with the proposed framework. It is worthy to note that, we do not show the result of the actual computation time, but only the actual time as a percentage of the expected time (${T}_\textrm{actual} / {T}_\textrm{total}$), namely the  $\textrm{Cost}_\textrm{effective}$ calculated from the equation \ref{costeffective}. The reason is that the actual computation time may vary with different hardware resources or cloud computing platforms, and we aim to achieve high cost-effectiveness by stopping the clustering process at an early point, regardless of the platforms and hardware settings. 


\begin{table*}
    \caption{Actual time (percentage) with different desired accuracies }
    \label{tab:actual time}
    \begin{tabular}{@{}llllll@{}}
    \toprule
    \multirow{2}{*}{\textbf{Algorithm}} & \multicolumn{5}{l}{\textbf{Desired accuracy}}                  \\ \cmidrule(l){2-6} 
                                &\textbf{85\%} &\textbf{90\%} & \textbf{95\%} & \textbf{99\%} & \textbf{99.9\%} \\ \cmidrule(r){1-6}
    \textit{FCM}          & 27.34\%      & 29.33\%       & 33.25\%       & 55.93\%       & 60.83\%         \\ \bottomrule
    \end{tabular}
\end{table*}

\subsection*{Discussion}

Fig.~\ref{fig:points} shows the relation between the accuracy and the change rate of objective function. It can be seen that the predicted change rate of objective function without anomalies is generally lower than the predicted value with anomalies, indicating that the proposed methods can achieve higher accuracy compared to the previous methods without anomaly detection algorithms when given the same desired accuracy.

\begin{figure}
    \centering
    \includegraphics[width=0.85\linewidth]{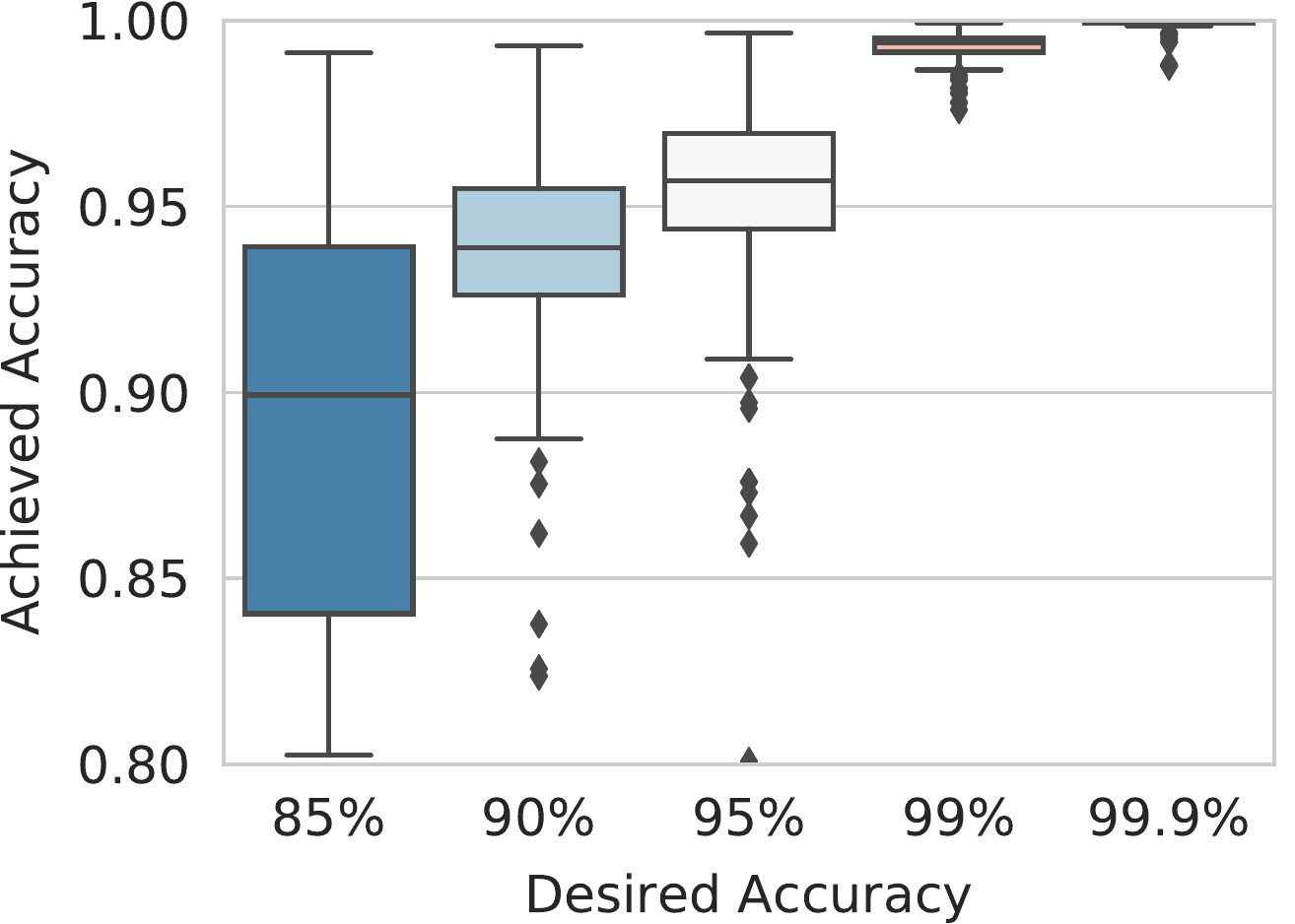}
    \caption{Boxplot of achieved accuracy over different desired accuracy}
    \label{fig:boxplot}
\end{figure}

Fig.~\ref{fig:boxplot} shows the boxplot between the achieved accuracy and the desired accuracy. It can be seen that the achieved accuracy is very close to the desired accuracy in different settings. The variation of accuracy becomes smaller with the increase of the desired accuracy, which proves the high performance of the proposed cost-effective land cover classification method

From the experiments, we have observed that the higher the desired accuracy, the longer the computation time and the less the time saved. by using the proposed approach, users can save more money with lower but sufficient accuracy (e.g., 90\%). For example, achieving 90\% accuracy needs only 29.33\% computation cost of 100\% accuracy. For the SpaceNet dataset, the training process is only computed once. The training process for 200 remote sensing images (using the FCM algorithm) took 6431.04 seconds and was only computed once. Taking the California land cover statistics as the instance for $423,970$ ${km}^2$ land, which needs around $2.567\times10^7$ partitioned remote sensing images ($438 \times 406$ pixels) with each covering a $16,520.74 m^2$ land. With the proposed approach, the saved computation time is approximately $162,035.31$ hours when the desired accuracy is 90\%. Based on Amazon EC2 pricing \cite{Amazon}, if we run m5.xlarge virtual machine instances (\$0.424 per hour), the cloud computation cost saved can be up to \$68,702.97 for California. Apparently, the cost in the training phase (\$0.378) is negligible to the whole computation cost. 

In real-world applications, the training phase is performed once and once completed, we can utilize the regression model many times. For instance, we can use the same regression model to carry out the whole United States land cover classification, which would save the computation cost up to \$1,593,490.18 in each single-use for the case of the desired accuracy of 90\%.

\section*{Conclusion}
Traditional land cover classification usually requires huge computational resources, and how to save computation costs in the cloud has become an increasingly important issue. For land cover classification, it is often not necessary to achieve the best accuracy all the time, usually no less than 85\% can be regarded as a reliable land cover classification. 
 
In this research, we proposed a generalized framework for cost-effective remote sensing classification. FCM algorithm was applied for clustering remote sensing images, with \textit{Rand Index} as the accuracy calculation method and Local Outlier Factors (LOF) as the anomaly detection algorithm. The Support Vector Regressor (SVR) was used to fit the relation between the change rate of objective function and accuracy. Extensive experimental results showed that given the desired accuracy (e.g., 85\%, 90\%, 95\%,99\%, 99.99\%), we can make the FCM clustering process on remote sensing images stop earlier and therefore save a huge amount of computation time. Also, the achieved accuracy (i.e., 89.15\%, 93.16\%, 95.07\%, 99.27\%, 99.92\%) are very close to yet above the desired accuracy. 

However, there are some threats to the validity of this research. One main threat is the representativeness of the data set used in the experiments. The real-world remote imagery dataset SpaceNet is used in our study. This dataset may have its own characteristics and may not comprehensively present all remote sensing datasets. Nevertheless, our framework is flexible and researchers can adjust the clustering algorithm, accuracy calculation method, anomaly detection algorithm, and regression model in different clustering scenarios (not limited to the SpaceNet datasets or even land cover classification problem) based on their own needs. They can also set the desired accuracy and then make the clustering algorithm stop early with sufficient accuracy to save much computation cost. Another threat is the representativeness of the experiment environment. We conduct experiments on the Microsoft Surface Laptop 4 with 64-bit Windows 10 enterprise, instead of using the EC2 virtual machines instances on Amazon cloud directly. The reason  we do not compare the performance of the proposed framework is that we aim to achieve high cost-effectiveness by stopping the clustering process at an early stop point, and the saved time by reducing the number of iterations is independent of the platform. In the future, the proposed framework can be easily ported to different cloud platforms such as AWS Lambda, EC2 and Azure. Therefore, the threats to the validity are minimal in this research.


In future research, we will focus on several aspects to improve our proposed framework. Firstly, we will compare the performance of different clustering algorithms using the proposed framework. Secondly, more remote sensing datasets will be explored to verify the robustness and the generality of the framework. Additionally, we will investigate methods to bound the achieved accuracy within a given error range.



%
%
%
%

\section*{Appendix}
The notations used in this research are shown in Table~\ref{tab:notation}.

\begin{table}[]
\caption{Table of notations in this research}
\label{tab:notation}
\begin{tabular}{@{}ll@{}}
\toprule
Notation   & Defination \\ \midrule
$m$ & The $m$th iteration during the clustering process      \\
$x$ & A d-dimensional feature point\\
$x_i$& The $i$th d-dimensional data point in cluster $x$         \\
$c_j$          & The d-dimensional center of the cluster $j$           \\
$u_{ij}$  & The degree of membership of $x_j$ in the cluster $j$          \\
$\mathcal{J}_m$& The objective function at the $m$th iteration \\
$P_i$ &  The $i$th partition during the clustering process\\
$Rand(P_1,P_2)$      & The \textit{Rand Index} of two partitions $P_1$ and $P_2$ \\
$\mathcal{L}_m$          & The predicted labels at the $m$th iteration \\
$r_m$          & The accuracy at the $m$ iteration          \\ 
$\Delta{\mathcal{J}_m}$          & The change rate of $\mathcal{J}_m$  \\ 
$\bar{r}$           & The desired accuracy (e.g., 85\%, 90\%, 95\%, etc.)          \\ 
 $\Delta \bar{\mathcal{J}}$   & The predicted objective function given the $\bar{r}$ \\ 
$\textrm{Cost}_\textrm{comp}$          & The computation cost in the cloud          \\ 
$\textrm{Price}_\textrm{unit}$         & The unit price          \\
${T}_\textrm{comp}    $     & Total computation time          \\ 
${T}_\textrm{train}  $        & The time taken in the training process\\ 
$T_\textrm{actual}      $    & The early-stop computation time in the clustering         \\ 
  
  ${T}_\textit{total}$ & The computation time for achieving 100\% accuracy\\ $\textrm{Cost}_\textrm{effective}$  & The cost-effectiveness percentage\\
\bottomrule
\end{tabular}
\end{table}


\begin{backmatter}

\section*{Acknowledgements}
This research is supported by the Beijing Institute of Technology and Swinburne University of Technology. We also sincerely thank the anonymous reviewers for this paper.

\section*{Funding}
This work is partly funded by the China Scholarship Council, National Key Research and Development Plan of China (2016YFB0502604, 2016YFC0803000), and Key Science and Technology Project of Beijing (Z171100005117002).


\section*{Availability of data and materials}
Not applicable.


\section*{Competing interests}
The authors declare that they have no competing interests.


\section*{Authors' contributions}
Dongwei Li designed the improved framework and conducted experiments under the supervision of Yun Yang, Qiang He and Shuliang Wang. All authors participated in the data analysis and manuscript writing. The authors read and approved the final manuscript.

\section*{Authors' information}

\textbf{Dongwei Li} received his M.Sc. degree in software engineering from Wuhan University, China, in 2010. He is working toward his Ph.D. degree at Beijing Institute of Technology, Beijing, China. He is a currently visiting researcher at Swinburne University of Technology, Australia. His research interests include data mining and cloud computing.

\textbf{Shuliang Wang} received his Ph.D. degree from Wuhan University, China, in 2002. He is a full professor at Beijing Institute of Technology, China. His research interests include spatial data mining, data field and big data.

\textbf{Qiang He} received his Ph.D. degree in information and communication technology from Swinburne University of Technology (SUT), Australia, in 2009. He is now a senior senior lecturer at Swinburne University of Technology. His research interests include software engineering, cloud computing and services computing.

\textbf{Yun Yang} received his Ph.D. degree from the University of Queensland, Australia in 1992. He is a full professor at Swinburne University of Technology, Australia. His research interests include distributed systems, cloud and edge computing, software technologies, workflow systems and service-oriented computing.

\bibliographystyle{bmc-mathphys} 
\bibliography{bmc_article}      

\end{backmatter}
\end{document}